# Analysis of Causality Issue in Near-field Superluminally Propagating Electromagnetic and Gravitational Fields

William D. Walker
Royal Institute of Technology, KTH-Visby
Department of Electrical Engineering
Cramérgatan 3, S-621 57 Visby, Sweden
bill@visby.kth.se

## 1   Introduction

A simple relativistic analysis is presented which shows that near-field superluminal longitudinal electric signals generated by an electric dipole cannot be used to violate Einstein causality by using the relativistic "sync shift" effect[a]. The analysis shows that because a signal has some time extent (i.e. after a signal is initiated one must wait at least one period for the frequency and amplitude information to be determined), an electric dipole can be used to transmit a signal backward in time, but not before the same signal was initiated, thereby prohibiting a user from changing the signal that was transmitted. Although the analysis is presented specifically for superluminal near-field longitudinal electric fields, the result would also apply for any type of signal that is superluminal in a spatial region less than one wavelength. Therefore the superluminal near-field transverse electric and magnetic fields, which are also generated by an electric dipole, cannot be used to violate causality using the "sync shift" effect, since they are known to be superluminal in a region less than one wavelength near the source. In addition, the gravitational fields generated by an oscillating mass, which are known to be superluminal in a region less than one wavelength near the source, cannot be used to violate causality using the "sync shift" effect.

## 2   Theoretical analysis

The following relativistic analysis looks at the consequences of transmitting a superluminal near-field longitudinal electric field from an amplitude-modulated electric dipole to a moving electron which then reflects the signal back to the source. The result is analysed to see if it is possible to transmit a signal backward in time so that the information can be used to change the same information that was transmitted, thereby violating causality.

---

[a] <u>Sync shift effect</u> – Term used to refer to the relativistic effect that enables superluminal signals to be transmitted backward in time. The Lorentz transformation for time change: $\Delta t' = \gamma [\Delta t - (v/c^2)\Delta x]$, where $\Delta t = L/w$ and $\Delta x = L$ (transmission distance), can become negative when the signal velocity (w) is much greater than the velocity (v) of a moving observer: $w > (c^2)/v$ [3].

A stationary electric dipole transmits a longitudinal electric field toward an electron which moves away at constant velocity (v) (ref. Figure 1).

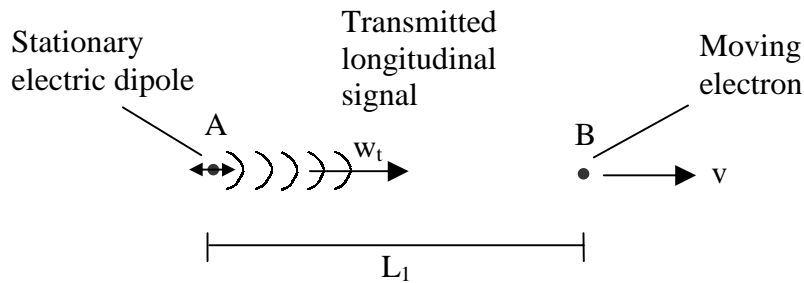

**Figure 1**: Stationary electric dipole (A) transmitting longitudinal signal to moving electron (B) at a distance ($L_1$) away from the stationary electric dipole.

The dipole and the moving electron are separated by distance ($L_1$). From the analysis presented in several papers by the author, the longitudinal electric field generated by an oscillating electric dipole is known to propagate (group speed) nearly instantaneously provided the propagation distance ($L_1$) is much less than one wavelength (~ $\lambda_T/10$) [1, 2, 3]. It is assumed that the signal to be transmitted is an amplitude-modulated signal with a modulation wavelength ($\lambda_T$) (ref. Figure 2). The beginning of the signal starts at time ($B_T$) and the end of the signal ends at time ($E_T$).

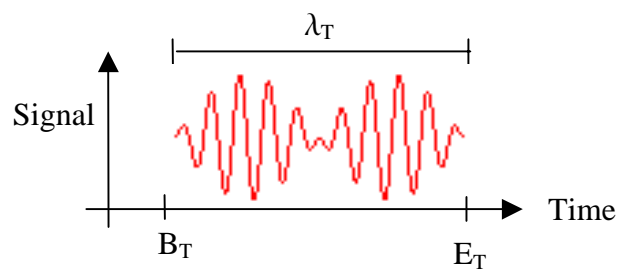

**Figure 2**: Diagram of an amplitude-modulated signal with modulation wavelength ($\lambda_T$). The modulated signal starts at time ($B_T$) and ends at time ($E_T$).

This process can be represented graphically using a Minkowski space-time diagram (ref. Figure 3).

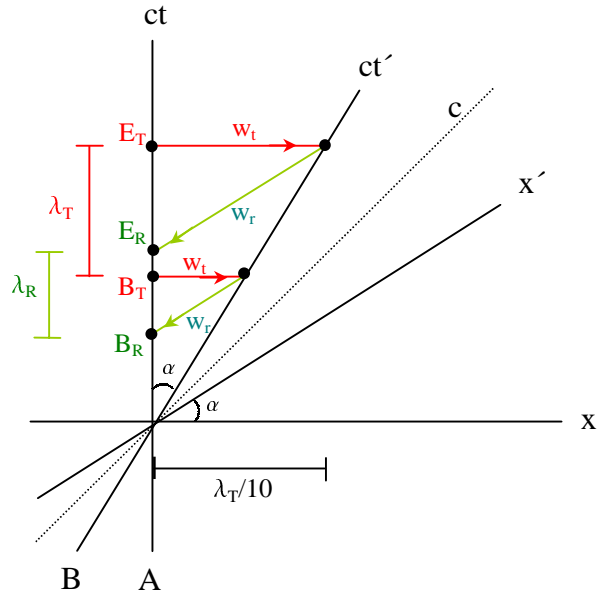

**Figure 3**: Space-time diagram of an instantaneous signal ($\lambda_T$) propagating from stationary point (A) to moving point (B), and then instantaneously reflecting back to point (A), resulting in signal ($\lambda_R$).

The diagram shows the space-time co-ordinates (x, ct) of the stationary electric dipole (A) superimposed with the moving electron (B) space-time co-ordinates (x´, ct´) [4]. The stationary electric dipole is timelike and is represented by the (ct) axis, and the moving electron (travelling with velocity v) is represented by the (ct´) axis. The signal (wavelength = $\lambda_T$) is transmitted nearly instantaneously ($w_t$) to the moving electron and is seen as a horizontal line intersecting the (ct) and (ct´) axes. Since the signal has some time extent (i.e. the signal has a period cT = c/f = $\lambda_T$), the end of the signal ($E_T$) will be transmitted some time ($\lambda_T$) after the beginning of the signal ($B_T$). The angle ($\alpha$) between the space-time co-ordinates is known to have the following relationship [Tan($\alpha$) = v/c]. This follows from the fact that the moving electron (ct´ axis) travelling with velocity (v) is related to the graph co-ordinates: v = x/t, therefore v/c = x/(ct) = Tan($\alpha$). If the longitudinal signal were then reflected ($w_r$) by the moving electron (B) toward the stationary electric dipole (A) (ref. Figure 4), it would propagate nearly instantaneously according to its co-ordinate system (i.e. parallel to the x´ axis).

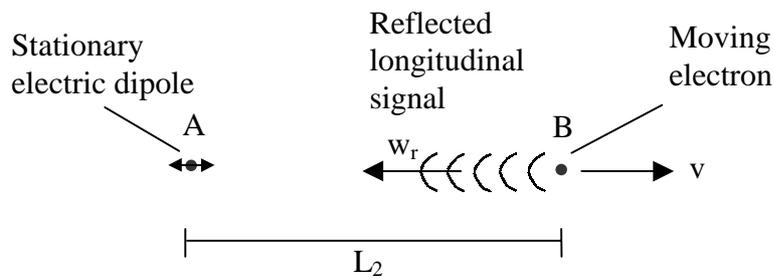

**Figure 4**: Moving electron (B) reflecting longitudinal signal back to stationary electric dipole (A) located a distance ($L_2$) away.

The beginning of the signal arrives at the electric dipole (A) at time ($B_R$) and the end of the signal arrives at the electric dipole (A) at time ($E_R$). Note that the largest separation distance between the dipole and the moving electron is ($\lambda_T/10$), thereby enabling the longitudinal signals to propagate nearly instantaneously. The time difference (c$\Delta$t) between the transmission of the end of the signal ($E_T$) and the return of the end of the signal ($E_R$) can then be calculated using simple trigonometric relationships (ref. Figure 5).

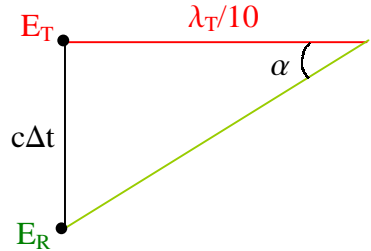

**Figure 5**: Trigonometric diagram relating the time difference (c$\Delta$t) between the transmission of the end of the signal ($E_T$) and the return of the end of the signal ($E_R$) to parameters ($\lambda_T$, $\alpha$).

From Figure 5 it can be seen that Tan($\alpha$) = (c$\Delta$t)/($\lambda_T/10$) and it is known that Tan($\alpha$) = v/c. Equating these two relations and solving for c$\Delta$t yields: c$\Delta$t = (v $\lambda_T$)/(10 c). This result can also be obtained by using Lorentz transformations. From the moving electron's perspective, the end of the signal ($E_T$) propagates backward in time: $\Delta$t´ = $\gamma$ [$\Delta$t - (v/c$^2$)$\Delta$x], where $\Delta$t = 0, $\Delta$x = $L_1$, and $\gamma$ = 1/Sqrt[1-(v/c)$^2$]. When the moving electron reflects the signal, from its perspective, the electric dipole moves with velocity (v) and sees a contracted distance $\Delta$x = $L_2$ = $L_1/\gamma$. The end of the signal ($E_R$) therefore arrives at the electric dipole (A) at time: $\Delta$t = $\gamma$ [0 - (v/c$^2$)$L_1/\gamma$], where $L_1$ = $\lambda_T/10$. Therefore one obtains the same solution as was obtained from the Minkowski space-time diagram: c$\Delta$t = -(v $\lambda_T$)/(10 c). Since the velocity of the moving electron (v) can be at most (c), then: c$\Delta$t < $\lambda_T/10$ < $\lambda_T$. This result indicates that although the signal can be transported backward in time ($\lambda_R$ arrives before $\lambda_T$), it cannot be transported before the same signal was initiated ($E_R$ > $B_T$), thus preserving causality. The signal transmitted backward in time cannot be used to change the same signal that was sent.

## 3  Conclusion

A simple relativistic analysis has been presented which shows that near-field superluminal longitudinal electric signals generated by an electric dipole cannot be used to violate Einstein causality with the relativistic "sync shift" effect. The analysis has shown that because a signal has some time extent (i.e. after a signal is initiated one must wait at least one period for the frequency and amplitude information to be determined), an electric dipole can be used to transmit a signal backward in time, but not before the same signal was initiated, thereby prohibiting a user from changing the signal that was transmitted.

It should be noted that in the model presented it was assumed that the signal information was contained in one period ($\lambda_T$) of an amplitude-modulated signal. In a real physical system several wavelengths may be required to encode and decode the information. One consequence of this is that it would reduce the spatial region over which the information travels superluminally [3]. In addition, it would also increase the effective signal wavelength ($\lambda_T$) used in the modelling of the problem, thus making it even more difficult for the signal to be transported before the same signal was initiated ($E_R \gg B_T$), and thus making it more difficult to violate causality.

Although the analysis is presented specifically for superluminal near-field longitudinal electric fields, the result would also apply for any type of signal that is superluminal in a region less than one wavelength. In order to violate causality a signal must be superluminal in a spatial region greater than at least one wavelength. Therefore, the superluminal near-field transverse electric and magnetic fields generated by an electric dipole cannot be used to violate causality using the "sync shift" effect, since these fields are also known to be superluminal in a region less than one wavelength near the source.

In addition, in reference [1] the gravitational fields generated by an oscillating mass are also found to be superluminal in a region less than one wavelength near the source. Analogous to the electric dipole and moving electron system, if a moving mass were to reflect the superluminal near-field gravitational fields generated by an amplitude-modulated mass, then the signal would arrive back in time at the transmitter mass, but not before the signal was initiated, thereby prohibiting a user from changing the same signal that was transmitted. It is therefore concluded that superluminal near-field gravitational fields cannot be used to violate causality using the "sync shift" effect.

In conclusion, contrary to Einstein's hypothesis that superluminal signals are incompatible with relativity theory [5], this paper has shown that superluminal signals, such as the superluminal near-field electromagnetic fields generated by an electric dipole or the superluminal near-field gravitational fields generated by a vibrating mass, are compatible with relativity because the spatial region in which they are superluminal is less that one wavelength. Although these superluminal fields can be used to transmit signals backward in time using the "sync shift" effect, the signals cannot be decoded before the same signals were initiated, thereby preserving causality.